\documentclass[iop,apj]{emulateapj}

\usepackage{graphicx}
\usepackage{verbatim}
\usepackage{parskip}
\usepackage{natbib}
\usepackage{hyperref,soul}

\received{January 19, 2016}
\accepted{March 10, 2016}
\journalinfo{Astronomical~Journal}

\newcommand{\Fermi}{{\it{}Fermi}\ }
\newcommand{\n}{\nodata}
\newcommand{\be}{\begin{itemize}}
\newcommand{\ee}{\end{itemize}}
\newcommand{\muasyr}{\hbox{$\; \mu{\rm as \ y}^{-1}\;$}}

\newlength\mystoreparindent
\newenvironment{myparindent}[1]{%
\setlength{\mystoreparindent}{\the\parindent}
\setlength{\parindent}{#1}
}{%
\setlength{\parindent}{\mystoreparindent}
}

\makeindex
\citeindextrue

\submitted{}
\shorttitle{MOJAVE. XIII. Parsec-Scale AGN Jet Kinematics 1994--2013}

\shortauthors{M. L. Lister et al.}

\begin{document}
\title{MOJAVE XIII. Parsec-Scale AGN Jet Kinematics Analysis Based on 19
  years of VLBA Observations at 15 GHz}

\author{M. L. Lister\altaffilmark{1},
M. F. Aller\altaffilmark{2},
H. D. Aller\altaffilmark{2},
D. C. Homan\altaffilmark{3},
K. I. Kellermann\altaffilmark{4},
Y. Y. Kovalev\altaffilmark{5,6},
\mbox{A. B. Pushkarev\altaffilmark{7,5}},
J. L. Richards\altaffilmark{1},
E. Ros\altaffilmark{6,8,9},
T. Savolainen\altaffilmark{10,6}
}

\altaffiltext{1}{
Department of Physics and Astronomy, Purdue University, 525 Northwestern Avenue,
West Lafayette, IN 47907, USA;
\email{mlister@purdue.edu}
}
\altaffiltext{2}{
Department of Astronomy, University of Michigan, 311 West Hall, 1085
S. University Avenue, Ann Arbor, MI 48109, USA;
}

\altaffiltext{3}{
Department of Physics, Denison University, Granville, OH 43023, USA;}

\altaffiltext{4}{
National Radio Astronomy Observatory, 520 Edgemont Road, Charlottesville, VA 22903, USA;
}

\altaffiltext{5}{
Astro Space Center of Lebedev Physical Institute,
Profsoyuznaya 84/32, 117997 Moscow, Russia;
}
\altaffiltext{6}{
Max-Planck-Institut f\"ur Radioastronomie, Auf dem H\"ugel 69,
53121 Bonn, Germany;
}

\altaffiltext{7}{
Crimean Astrophysical Observatory, 98409 Nauchny, Crimea, Russia;}

\altaffiltext{8}{
Observatori Astron\`omic, Universitat de Val\`encia,
  Parc Cient\'{\i}fic, C. Catedr\'atico Jos\'e Beltr\'an 2, E-46980
  Paterna, Val\`encia, Spain}

\altaffiltext{9}{
Departament d'Astronomia i Astrof\'{\i}sica,
  Universitat de Val\`encia, C. Dr. Moliner 50, E-46100 Burjassot,
  Val\`encia, Spain}

\altaffiltext{10}{Aalto University Mets\"ahovi Radio Observatory, Mets\"ahovintie 114, FI-02540 Kylm\"al\"a, Finland}

\begin{abstract}

  We present 1625 new 15 GHz (2 cm) VLBA images of 295 jets associated
  with active galactic nuclei (AGNs) from the MOJAVE and 2 cm VLBA
  surveys, spanning observations between 1994 Aug 31 and 2013 Aug 20.
  For 274 AGNs with at least 5 VLBA epochs, we have analyzed the
  kinematics of 961 individual bright features in their parsec-scale
  jets. A total of 122 of these jets have not been previously analyzed
  by the MOJAVE program. In the case of 451 jet features that had at
  least 10 epochs, we also examined their kinematics for possible
  accelerations.  At least half of the well-sampled features have
  non-radial and/or accelerating trajectories, indicating that
  non-ballistic motion is common in AGN jets. Since it is impossible
  to extrapolate any accelerations that occurred before our monitoring
  period, we could only determine reliable ejection dates for $\sim
  24\%$ of those features that had significant proper motions.  The
  distribution of maximum apparent jet speeds in all 295 AGNs measured
  by our program to date is peaked below $5c$, with very few jets with
  apparent speeds above $30c$.  The fastest speed in our survey is
  $\sim 50c$, measured in the jet of the quasar PKS 0805$-$07, and is
  indicative of a maximum jet Lorentz factor of $\sim 50$ in the
  parent population. An envelope in the maximum jet speed versus
  redshift distribution of our sample provides additional evidence of
  this upper limit to the speeds of radio-emitting regions in
  parsec-scale AGN jets. The {\it Fermi} LAT-detected gamma-ray AGNs
  in our sample have, on average, higher jet speeds than non
  LAT-detected AGNs, indicating a strong correlation between pc-scale
  jet speed and gamma-ray Doppler boosting factor.  We have identified
  11 moderate-redshift ($z < 0.35$) AGNs with fast apparent speeds ($>
  10c$) that are strong candidates for future TeV gamma-ray
  detection.  Of the five gamma-ray loud narrow-lined Seyfert I AGNs
  in our sample, three show highly superluminal jet motions, while the
  others have sub-luminal speeds.  This indicates that some
  narrow-lined Seyfert I AGNs possess powerful jets with Lorentz
  factors in excess of 10, and viewing angles less than $10\arcdeg$,
  consistent with those of typical BL Lac objects and flat-spectrum
  radio quasars.
\end{abstract}
\keywords{
galaxies: active ---
galaxies: jets ---
radio continuum: galaxies ---
quasars: general ---
BL Lacertae objects: general
} 
 

\section{INTRODUCTION} 
\label{s:intro}

Very long baseline interferometry (VLBI) provides the highest
resolution images in astronomy, and is invaluable for studying the
structure and evolution of non-thermal radio sources. The powerful
jetted outflows associated with active galactic nuclei (AGNs) are a
natural target for VLBI, since they are exceedingly compact, and
dominate the extragalactic sky at radio and high energy wavelengths.
Because of their relativistic flow speeds, AGN jets are affected by
beaming and aberration effects that can significantly enhance their
apparent luminosity and variability, and create the illusion of
superluminal motion in the sky plane \citep{1977Natur.267..211B}. AGNs
of the rare blazar class represent cases where a high-Lorentz factor
jet is pointed nearly directly at us, thereby maximizing these
relativistic Doppler effects.

\begin{deluxetable*}{llllccll}
\centering
\tablecolumns{8} 
\tabletypesize{\scriptsize} 
\tablewidth{0pt}  
\tablecaption{\label{gentable} AGN Properties}  
\tablehead{ \colhead{B1950}&\colhead{J2000}  & \colhead {Alias} &\colhead{Gamma-ray Assoc.} &  
  \colhead{Opt.} & \colhead{1.5 Jy} &\colhead{$z$}&\colhead{Reference}  \\ 
\colhead{(1)} & \colhead{(2)} & \colhead{(3)} & \colhead{(4)} & \colhead{(5)} & 
 \colhead{(6)} & \colhead{(7)}& \colhead{(8)} } 
\startdata 
0003+380\tablenotemark{a} &J0005+3820 &  S4 0003+38&3FGL J0006.4+3825& Q & \n   & 0.229 & \cite{1994AAS..103..349S} \\ 
0003$-$066\tablenotemark{a} &J0006$-$0623 &  NRAO 005&\n& B & Y   & 0.3467 & \cite{2005PASA...22..277J} \\ 
0006+061\tablenotemark{a} &J0009+0628 &  CRATES J0009+0628&3FGL J0009.1+0630\tablenotemark{e}& B & \n   & \n & \cite{2012AA...538A..26R} \\ 
0007+106\tablenotemark{a} &J0010+1058 &  III Zw 2&\n& G & Y   & 0.0893 & \cite{1970ApJ...160..405S} \\ 
0010+405\tablenotemark{a} &J0013+4051 &  4C +40.01&\n& Q & \n   & 0.256 & \cite{1992ApJS...81....1T} \\ 
0015$-$054\tablenotemark{a} &J0017$-$0512 &  PMN J0017$-$0512&3FGL J0017.6$-$0512& Q & \n   & 0.226 & \cite{2012ApJ...748...49S} \\ 
0016+731\tablenotemark{a} &J0019+7327 &  S5 0016+73&\n& Q & Y   & 1.781 & \cite{1986AJ.....91..494L} \\ 
0027+056\tablenotemark{a} &J0029+0554 &  PKS 0027+056&\n& Q & \n   & 1.317 & \cite{1999AJ....117...40S} \\ 
0026+346\tablenotemark{a} &J0029+3456 &  B2 0026+34&\n& G & \n   & 0.517 & \cite{2002AJ....124..662Z} \\ 
0035+413\tablenotemark{a} &J0038+4137 &  B3 0035+413&\n& Q & \n   & 1.353 & \cite{1993AAS..101..521S} \\ 
0048$-$097\tablenotemark{a} &J0050$-$0929 &  PKS 0048$-$09&3FGL J0050.6$-$0929& B & Y   & 0.635 & \cite{2012AA...543A.116L}
\enddata 
\tablecomments{This is a table stub, the full version is available as an ancillary file.  Columns are as follows: 
(1) B1950 name, 
(2) J2000 name, 
(3) other name,
(4) gamma-ray association name,
(5) optical classification, where B = BL Lac, Q = quasar, G = radio galaxy, N = narrow-lined Seyfert 1, and U = unidentified, 
(6) MOJAVE 1.5 Jy sample membership flag,
(7) redshift,
(8) reference for redshift and/or optical classification.
}

\tablenotetext{a}{Jet kinematics analyzed in this paper, based on data up to 2013 Aug 20.}
\tablenotetext{b}{Jet kinematics analyzed by \cite{MOJAVE_X}, based on data up to 2011 May 1.}
\tablenotetext{c}{Known TeV emitter (\url{http://tevcat.uchicago.edu}).}
\tablenotetext{d}{{\it Fermi} LAT detection reported by \cite{2014ATel.5838....1C}}
\tablenotetext{e}{One of two AGN associations listed for this 3FGL source in \cite{3LAC} (see Appendix)}
\end{deluxetable*}

Starting with the 2 cm VLBA survey in 1994
\citep{1998AJ....115.1295K}, and continuing with the MOJAVE
(Monitoring of Jets in Active Galactic Nuclei With VLBA Experiments)
program \citep{MOJAVE_V}, we have carried out multi-epoch VLBA
observations of several hundred of the brightest, most compact radio
sources in the northern sky.  These are predominantly blazars, due to
the selection biases associated with relativistic beaming
\citep{BlandfordKonigl79}. 

In this paper, we present 1625 VLBA 15 GHz contour maps of 295 AGNs
for epochs between 1994 Aug 31 and 2013 Aug 20 that have not
previously appeared in any paper from the MOJAVE \citep{MOJAVE_I,
  MOJAVE_V, MOJAVE_X} or 2 cm VLBA surveys
\citep{1998AJ....115.1295K,2cmpaperII}.  These AGNs (see
Table~\ref{gentable}) are from one or more of the following: the
MOJAVE low-luminosity sample \citep{MOJAVE_X}, the complete flux
density-limited MOJAVE 1.5 Jy sample \citep{2015ApJ...810L...9L}, the
VLBA 2 cm survey \citep{1998AJ....115.1295K}, the 3rd EGRET gamma-ray
catalog \citep{Hartman99}, or the 3FGL {\it Fermi} gamma-ray catalog
\citep{3FGL}.  Also included are some AGNs that were originally
candidates for these samples, but did not meet the final selection
criteria.

A major aspect of our program is to analyze these multi-epoch VLBA
observations in order to investigate the evolution of pc-scale AGN
jets, which we have reported on in a series of papers
\citep{2004ApJ...609..539K,MOJAVE_VI, MOJAVE_VII,
  MOJAVE_X,MOJAVE_XII}. Most AGN jets have bright, compact
radio-emitting features that move outward at apparent superluminal
speeds.  These features often change both their speed and direction,
suggesting that their kinematics are strongly affected by both
hydrodynamic and MHD plasma effects (e.g.,
\citealt{2003ApJ...589L...9H, 2015ApJ...803....3C}). In our most
recent analysis \citep{MOJAVE_X, MOJAVE_XII}, we studied the
kinematics of 200 AGN jets, for which at least 5 VLBA epochs were
obtained between 1994 Aug 31 to 2011 May 1.  These AGNs were drawn
from the MOJAVE 1.5 Jy radio flux density-limited and 1FM {\Fermi}
gamma-ray selected samples, described by \cite{2015ApJ...810L...9L}
and \cite{2011ApJ...742...27L}, respectively.  Here we present time
lapse movies and new kinematics analyses of 173 AGNs drawn from these
two samples, and 101 AGNs drawn from the other surveys described in
the previous paragraph, using VLBA data obtained between 1994 Aug 31
and 2013 Aug 20.  A total of 122 of these jets have not previously had
their kinematics analyzed by the MOJAVE program.  We have excluded
from our kinematics analysis 48 MOJAVE AGNs we previously analyzed in
\cite{MOJAVE_X}, since we are currently gathering additional VLBA
epochs on them.

The overall layout of the paper is as follows. In Section~\ref{data}
we describe our observational data and our method of modeling the
individual jet features, and in Section 3 we discuss their kinematic
properties and overall trends in the data. We summarize our findings
in Section~\ref{conclusions}. We adopt a cosmology with $\Omega_m =
0.27$, $\Omega_\Lambda = 0.73$ and $H_o = 71 \; \mathrm{km\; s^{-1} \;
  Mpc^{-1}}$ \citep{Komatsu09}. We refer to the AGNs throughout using
either B1950 nomenclature or commonly-used aliases, as listed in
Table~\ref{gentable}.

\section{OBSERVATIONAL DATA AND ANALYSIS}\label{data}

\begin{figure*}
\centering
\includegraphics[trim=0cm 0cm 0cm 0cm,height=0.93\textheight]{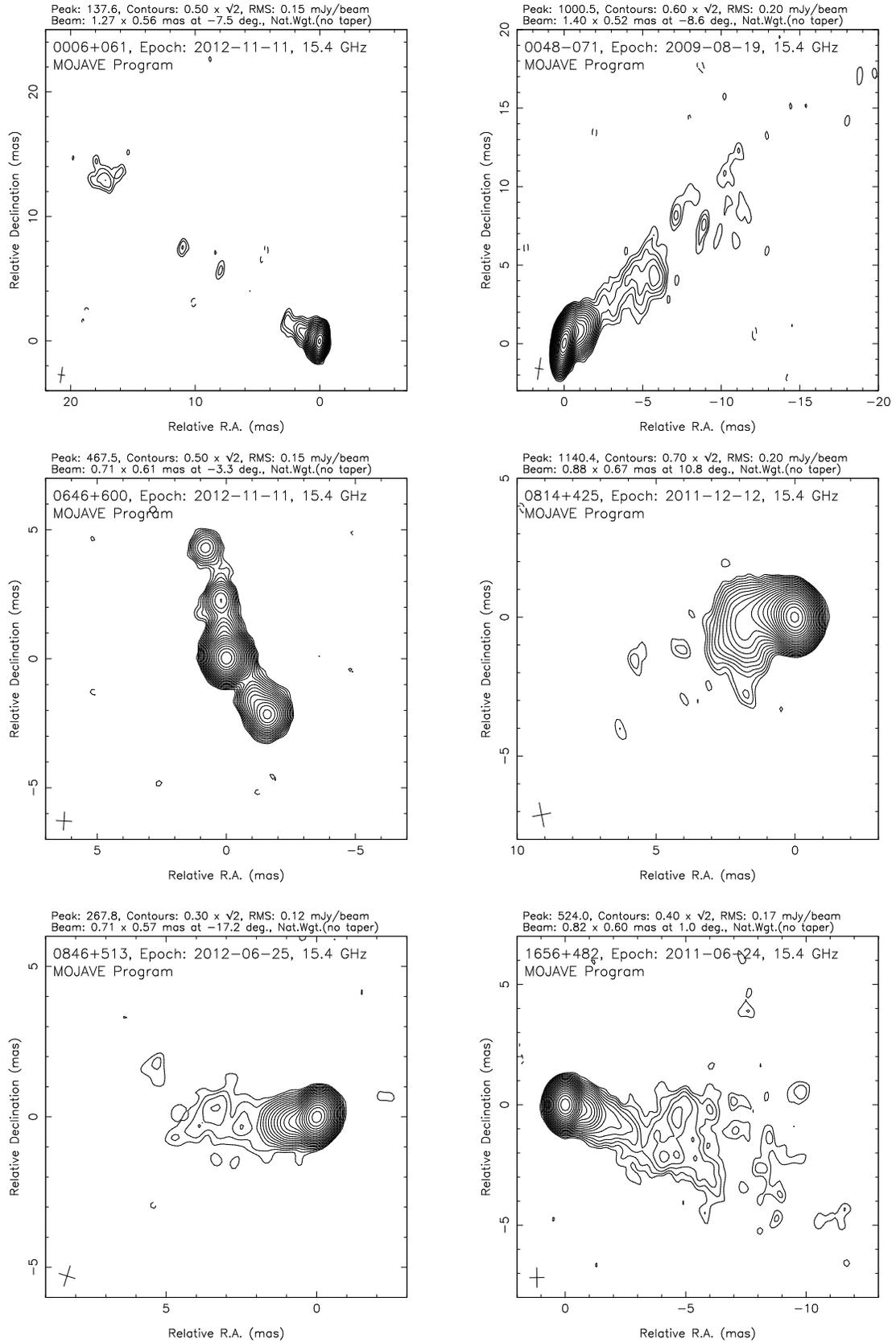}
\caption{\label{images} 
  Naturally-weighted 15~GHz total
  intensity VLBA contour images of individual epoch observations of
  the MOJAVE AGN sample. The contours are in successive powers of
  $\sqrt{2}$ times the base contour level in mJy per beam, as listed
  in Table~\ref{maptable} and at the top of each panel. The FWHM
  restoring beam dimensions are indicated as a cross in the lower left
  corner. Because of self-calibration, in some cases the origin may be
  coincident with the brightest feature in the image, rather than the
  putative core feature listed in Table~\ref{gaussiantable}. This is
  a figure stub. All images are available online at
  \url{http://www.astro.purdue.edu/MOJAVE/allsources.html}.}
\end{figure*}

\subsection{Contour Maps}

We processed the VLBA interferometric visibility data and produced
contour maps (Figure~\ref{images}) using the AIPS and Difmap
\citep{difmap} software packages, using the methods described by
\cite{MOJAVE_V, MOJAVE_X}.  The raw correlated data from the MOJAVE
program are available immediately after correlation from the NRAO data
archive\footnote{\url{http://archive.nrao.edu}}, and our processed data and
maps are publicly available on our project
website\footnote{\url{http://www.astro.purdue.edu/MOJAVE}} within a few
weeks after correlation. We list the parameters of the contour maps in
Table~\ref{maptable}.  The FWHM dimensions of the VLBA interferometric
restoring beam at 15 GHz vary with declination, but are on the order
of 1 mas (N-S) $\times$ 0.5 mas (E-W), which corresponds to a linear
resolution of a few parsecs (projected) at the median redshift ($z
\simeq 1$) of our sample.  Column 4 of Table~\ref{maptable} lists the
VLBA project code for each observation, along with an indicator of
whether it is from the MOJAVE program, the VLBA 2 cm Survey, or the
NRAO data archive.  For the latter, we considered only archival epochs
with at least 4 individual scans that spanned a reasonably wide range
of hour angle, and that included at least 8 VLBA antennas. The VLBA 2
cm Survey observations (1994--2002) consisted of approximately one
hour integrations on each AGN, broken up into several scans separated
in hour angle to improve the interferometric coverage. A similar
observing method was used in the full polarization MOJAVE observations
from 2002 May to 2007 September (VLBA codes BL111, BL123, BL137, and
BL149), and is described by \cite{MOJAVE_I}.  During 2006 (VLBA code
BL137), the 15 GHz integration times were shortened by a factor of
$\sim 3$ to accommodate interleaved scans at three other observing
frequencies (8.1, 8.4, 12.1 GHz). The latter were published by
\cite{MOJAVE_VIII}, \cite{MOJAVE_IX}, and \cite{MOJAVE_XI}. The MOJAVE
and 2 cm Survey observations were originally recorded at a data rate
of 128 Mbps, which was increased to 256 Mbps for the epochs from 2007
July 3 to 2008 Sept 12 inclusive, and 512 Mbps thereafter.  Beginning
with the 2007 Jan 6 epoch, we increased the number of AGNs observed in
each 24 hour MOJAVE session from 18 to 25 AGNs, and to 30 AGNs on 2009
Feb 25.

\begin{deluxetable*}{llllccrrrccc} 
\tablecolumns{12} 
\tabletypesize{\scriptsize} 
\tablewidth{0pt}  
\tablecaption{\label{maptable}Summary of 15 GHz Image Parameters}  
\tablehead{ & &  & \colhead{VLBA} & \colhead{Freq.} & 
\colhead{$\mathrm{B_{maj}}$} &\colhead{$\mathrm{B_{min}}$} & \colhead{$\mathrm{B_{pa}}$} &  
\colhead{$\mathrm{I_{tot}}$} &  \colhead{rms}  &  \colhead{$\mathrm{I_{base}}$} & \colhead{Fig.} \\ 
\colhead{Source} & \colhead{Alias} & \colhead {Epoch} & \colhead{Code} & \colhead{(GHz)} & 
\colhead{(mas)} &\colhead{(mas)} & \colhead{(\arcdeg)} &  
\colhead{(Jy)} & \colhead{(mJy}  &  \colhead{(mJy} & \colhead{Num.} \\ 
 & & & & & & & & & \colhead{bm$^{-1}$)} & \colhead{bm$^{-1}$)} \\ 
\colhead{(1)} & \colhead{(2)} & \colhead{(3)} & \colhead{(4)} &  
\colhead{(5)} & \colhead{(6)} & \colhead{(7)} & \colhead{(8)} & \colhead{(9)}& \colhead{(10)} &  \colhead{(11)} &  \colhead{(12)} } 
\startdata 
0003+380 & S4 0003+38  & 2011 Jun 6 & BL149DJ\tablenotemark{a} & 15.4 & 0.91 & 0.54 & $-$10 & 0.605 & 0.18 & 0.50 & \ref{images}.1   \\ 
 &   & 2013 Aug 12 & BL178BH\tablenotemark{a} & 15.4 & 0.84 & 0.53 & $-$4 & 0.667 & 0.20 & 0.50 & \ref{images}.2   \\ 
0003$-$066 & NRAO 005  & 2011 Jun 24 & BL149DL\tablenotemark{a} & 15.4 & 1.42 & 0.54 & $-$5 & 2.076 & 0.18 & 1.30 & \ref{images}.3   \\ 
 &   & 2012 Mar 4 & BL178AH\tablenotemark{a} & 15.4 & 1.31 & 0.53 & $-$3 & 2.132 & 0.17 & 0.60 & \ref{images}.4   \\ 
 &   & 2012 Nov 2 & BL178AR\tablenotemark{a} & 15.4 & 1.37 & 0.53 & $-$7 & 2.273 & 0.16 & 0.60 & \ref{images}.5   \\ 
0006+061 & CRATES J0009+0628  & 2011 Dec 29 & BL178AD\tablenotemark{a} & 15.4 & 1.32 & 0.65 & 8 & 0.209 & 0.22 & 0.60 & \ref{images}.6   \\ 
 &   & 2012 Jul 12 & BL178AM\tablenotemark{a} & 15.4 & 1.28 & 0.52 & $-$13 & 0.200 & 0.25 & 0.70 & \ref{images}.7   \\ 
 &   & 2012 Nov 11 & BL178AS\tablenotemark{a} & 15.4 & 1.27 & 0.56 & $-$8 & 0.182 & 0.15 & 0.50 & \ref{images}.8   \\ 
 &   & 2012 Dec 23 & BL178AX\tablenotemark{a} & 15.4 & 1.34 & 0.53 & $-$3 & 0.184 & 0.20 & 0.60 & \ref{images}.9   \\ 
 &   & 2013 Jun 2 & BL178BD\tablenotemark{a} & 15.4 & 1.32 & 0.56 & $-$1 & 0.162 & 0.23 & 0.70 & \ref{images}.10   \\ 
0007+106 & III Zw 2  & 1998 Feb 16 & BF039\tablenotemark{c} & 15.4 & 1.14 & 0.51 & $-$8 & 0.784 & 0.16 & 0.50 & \ref{images}.11   \\ 
 &   & 1998 Jun 13 & BF039B\tablenotemark{c} & 15.4 & 1.13 & 0.53 & $-$5 & 0.982 & 0.15 & 0.40 & \ref{images}.12   \\ 
 &   & 1998 Sep 14 & BG044\tablenotemark{c} & 15.4 & 1.11 & 0.49 & $-$4 & 1.202 & 0.18 & 0.70 & \ref{images}.13   \\ 
 &   & 2011 May 26 & BL149DI\tablenotemark{a} & 15.4 & 1.18 & 0.54 & $-$5 & 0.169 & 0.19 & 0.60 & \ref{images}.14
\enddata 
\tablecomments{This is a table stub, the full version is available as an ancillary file. 
Columns are as follows: (1) B1950 name, (2) other name, (3) date of VLBA observation, (4) VLBA experiment code, (5) observing frequency (GHz), (6) FWHM major axis of restoring beam (milliarcseconds), (7) FWHM minor axis of restoring beam (milliarcseconds), (8) position angle of major axis of restoring beam (degrees), (9) total I flux density (Jy),  (10) rms noise level of image (mJy per beam), (11) lowest I contour (mJy per beam), (12) figure number.}
\tablenotetext{a}{Full polarization MOJAVE epoch}
\tablenotetext{b}{2 cm VLBA Survey epoch}
\tablenotetext{c}{NRAO archive epoch}
\end{deluxetable*} 

\subsection{Time-lapse Movies}

\begin{figure}
\centering
\includegraphics[trim=2.5cm 0cm 3cm 1.5cm,clip,width=0.4\textwidth]{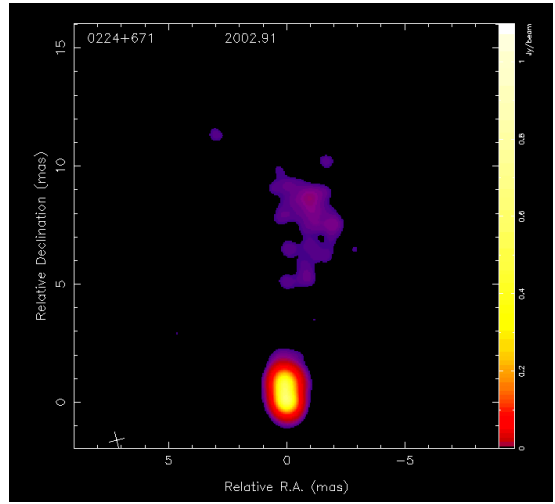}
  \caption{\label{movies} 
    Linearly interpolated time-lapse
    movie made from the multi-epoch VLBA images for 4C\,+67.05.  Each
    epoch is restored with a median beam, whose FWHM dimensions are
    indicated by the cross in the lower left corner. The false-color
    corresponds to radio flux density in units of Jy per beam
    (indicated by scale bar on the right side of the frames). One year
    of calendar time corresponds to 2.5~seconds in the movies. This
    movie is available at
    \url{http://www.astro.purdue.edu/MOJAVE/animated/0224+671.i.mpg}.}
\end{figure}

\begin{figure}
\centering
\includegraphics[trim=2.5cm 0cm 3cm 1.5cm,clip,width=0.4\textwidth]{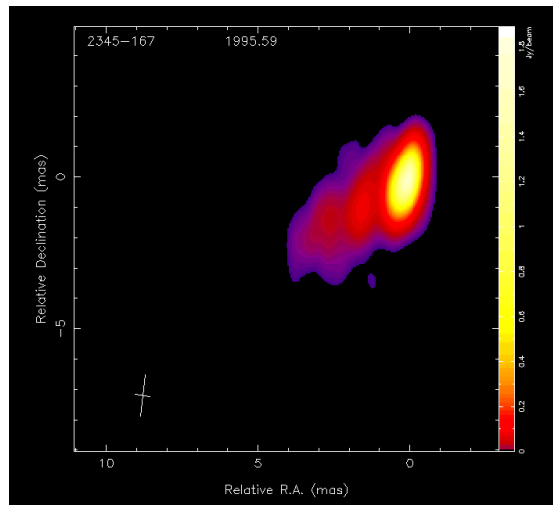}
  \caption{\label{movies2} Same as Figure~\ref{movies},
    for the quasar PKS~2345$-$16. This movie is available at
    \url{http://www.astro.purdue.edu/MOJAVE/animated/2345-167.i.mpg}.}
\end{figure}

In Figures~\ref{movies} and \ref{movies2} we show linearly
interpolated time-lapse MPEG movies of the multi-epoch VLBA images for
two selected AGNs (4C +67.05 and PKS 2345$-$16).  Movies for the other
AGNs can be found on the MOJAVE website.  We constructed the movies
using a two-point linear interpolation across each successive epoch,
treating each map pixel independently.  Prior to interpolation, we
restored all of the epoch maps to a scale of 0.05 milliarcseconds per
pixel using identical median beam dimensions that were based on the
full set of naturally weighted VLBA epochs available for that AGN.  We did
not interpolate across any time gaps larger than 4 years, leaving
these periods instead as blank frames in the movie.  The false-color
corresponds to radio flux density in units of Jy per beam (indicated
by scale bar on the right side of the frames). One year of calendar
time corresponds to 2.5 seconds of run time in the movies.

\subsection{Gaussian Modeling}

\begin{deluxetable*}{lclcrrcrcc} 
\tablecolumns{10} 
\tabletypesize{\scriptsize} 
\tablewidth{0pt}  
\tablecaption{\label{gaussiantable}Fitted Jet Features}  
\tablehead{\colhead{} & \colhead {} &   \colhead {} & 
 \colhead{I} & \colhead{r} &\colhead{P.A.} & \colhead{Maj.} & 
\colhead{} &\colhead{Maj. P.A.}   \\  
\colhead{Source} & \colhead {I.D.} &  \colhead {Epoch} & 
\colhead{(Jy)} & \colhead{(mas)} &\colhead{(\arcdeg)} & \colhead{(mas)} & 
\colhead{Ratio} &\colhead{(\arcdeg)}&\colhead{Robust?}   \\  
\colhead{(1)} & \colhead{(2)} & \colhead{(3)} & \colhead{(4)} &  
\colhead{(5)} & \colhead{(6)} & \colhead{(7)} & \colhead{(8)} & 
 \colhead{(9)} &  \colhead{(10)}} 
\startdata 
0003+380  & 0& 2006 Mar 9  & 0.489  & 0.04 & 290.7 & 0.23 & 0.33 & 292 & Y\\ 
  & 1&   & 0.007  & 3.98 & 121.8 & 0.72 & 1.00 & \n & Y\\ 
  & 2&   & 0.042  & 1.25 & 110.5 & 0.51 & 1.00 & \n & Y\\ 
  & 6&   & 0.104  & 0.28 & 114.6 & 0.27 & 1.00 & \n & Y\\ 
  & 7&   & 0.003  & 2.31 & 119.3 & \n & \n & \n & N\\ 
  & 0& 2006 Dec 1  & 0.320  & 0.10 & 308.1 & 0.25 & 0.29 & 295 & Y\\ 
  & 1&   & 0.005  & 3.65 & 120.8 & 1.63 & 1.00 & \n & Y\\ 
  & 2&   & 0.021  & 1.56 & 111.0 & 0.25 & 1.00 & \n & Y\\ 
  & 5&   & 0.023  & 0.75 & 116.2 & 0.32 & 1.00 & \n & Y\\ 
  & 6&   & 0.145  & 0.45 & 116.3 & 0.05 & 1.00 & \n & Y 
\enddata 
\tablenotetext{a}{Individual feature epoch not used in kinematic fits.}
\tablecomments{This is a table stub, the full version is available as an ancillary file. 
Columns are as follows: (1) B1950 name, (2) feature identification number (zero indicates core feature), (3) observation epoch, (4) flux density in Jy, (5) position offset from the core feature (or map center for the core feature entries) in milliarcseconds, (6) position angle with respect to the core feature (or map center for the core feature entries) in degrees,  (7) FWHM major axis of fitted Gaussian in milliarcseconds, (8) axial ratio of fitted Gaussian, (9) major axis position angle of fitted Gaussian in degrees, (10) robust feature flag.}
\end{deluxetable*} 

We modeled the $(u,v)$ visibility data in Difmap using a set of
Gaussian features, which we list in Table~\ref{gaussiantable}.  In
some instances, it was not possible to robustly cross-identify the
same features in a jet from one epoch to the next. We indicate those
features with robust cross-identifications across at least 5 epochs in
column 10 of Table~\ref{gaussiantable}. For the non-robust features,
we caution that the assignment of the same identification number
across epochs does not necessarily indicate a reliable
cross-identification. 

Based on previous analysis \citep{MOJAVE_VI}, we estimate the typical
uncertainties in the Gaussian centroid positions to be $\sim 20$\% of
the FWHM beam dimensions. For isolated bright and compact features,
the positional errors are smaller by approximately a factor of two.
We estimate the formal errors on the feature sizes to be roughly twice the
positional error, according to \cite{1999ASPC..180..301F}.  The flux
density accuracies are approximately 5\% (see Appendix A of
\citealt{2002ApJ...568...99H}), but can be significantly larger for
features located very close to one another. Also, at some epochs which
had antenna dropouts, the fit errors of some features are much larger.
We do not use the latter in our kinematics analysis, and indicate them by
flags in Table~\ref{gaussiantable}.

\subsection{Jet Kinematics Analysis}\label{kinematics}

As in \cite{MOJAVE_VI}, \cite{MOJAVE_VII}, \cite{MOJAVE_X}, and
\cite{MOJAVE_XII}, we analyzed the kinematics of the robust Gaussian
jet features in our sample using two methods. In the first method we
assumed a non-accelerating, two-dimensional (right ascension and
declination) vector fit to the feature position over time, referenced
to the core feature (which we presumed to be stationary).  The latter
represents the region in our VLBA images near the base of the jet
where the emission becomes optically thick at 15 GHz.  For the
features that had measurements at 10 or more epochs, our second method
employed a constant acceleration model that yielded kinematic fit
quantities at a (midpoint) reference date located exactly halfway
between the first and last VLBA observation dates for that particular
AGN (see column 9 of Table~\ref{acceltable}). The results of these
analyses are listed separately in Table~\ref{velocitytable} and
Table~\ref{acceltable}, respectively.  For completeness, these tables
also include kinematic fit parameters for 48 AGNs (indicated by flags
in column 1) that we analyzed in \cite{MOJAVE_X}, but not in this
paper.  We note that for features that show significant ($\ge3\sigma$)
accelerations, the acceleration-fit speed and other parameters in
Table~\ref{acceltable} provide a better description of their motions,
hence for these features we will use the acceleration fit quantities
in subsequent statistical analyses and discussion of their kinematics.

\begin{deluxetable*}{lcrrrrrrrrrrrr} 
\tablecolumns{14} 
\tabletypesize{\scriptsize} 
\tablewidth{0pt}  
\tablecaption{\label{velocitytable}Vector Motion Fit Properties of Jet Features}  
\tablehead{\colhead{} & \colhead {} &   \colhead {} & 
\colhead{$\langle S\rangle$}  &\colhead{$\langle R\rangle$} &\colhead{$\langle d_{\mathrm{proj}}\rangle$} & \colhead{$\langle\vartheta\rangle$} & 
 \colhead{$\phi$}&   \colhead{$ |\langle\vartheta\rangle - \phi|$}  &\colhead{$\mu$}  & \colhead{$\beta_{app}$} &  &   &    \\  
\colhead{Source} & \colhead {I.D.} &  \colhead {N} & 
\colhead{(mJy)} &\colhead{(mas)} & \colhead{(pc)} & \colhead{(deg)}   & 
\colhead{(deg)}& \colhead{(deg)} &\colhead{($\mu$as y$^{-1})$}& \colhead{($c$)}  &\colhead{$t_{ej}$}  & \colhead{$t_\alpha$}& \colhead{$t_\delta$}  \\  
\colhead{(1)} & \colhead{(2)} & \colhead{(3)} & \colhead{(4)} &  
\colhead{(5)} & \colhead{(6)} & \colhead{(7)} & \colhead{(8)} & 
 \colhead{(9)}& \colhead{(10)}&  
\colhead{(11)} & \colhead{(12)} & \colhead{(13)} & \colhead{(14)}   } 
\startdata 
0003+380  & 1 & 9  & 5 &4.24&  15.40& $ 120.5$ &
 97$\pm$16 & 24$\pm$16 & 157$\pm$42 & 2.29$\pm$0.61 &  \n &1985.09 &1888.98 \\ 
  & 2 & 7  & 17 &1.81&  6.57& $ 112.6$ &
 119.7$\pm$2.6 & 7.1$\pm$2.7 & 319$\pm$22 & 4.64$\pm$0.32 &  \n &2001.68 &2003.32 \\ 
  & 4 & 6  & 17 &1.25&  4.54& $ 114.8$ &
 208$\pm$12 & 93$\pm$12\tablenotemark{c} & 36.3$\pm$9.2 & 0.53$\pm$0.13 &  \n &2075.66 &1992.93 \\ 
  & 5 & 9  & 39 &0.76&  2.77& $ 117.5$ &
 331$\pm$119 & 146$\pm$119 & 4.8$\pm$7.2\tablenotemark{e} & 0.07$\pm$0.10 &  \n &2300.95 &2092.52 \\ 
  & 6 & 10  & 98 &0.39&  1.43& $ 115.4$ &
 335$\pm$41 & 141$\pm$41 & 12.7$\pm$8.3\tablenotemark{e} & 0.19$\pm$0.12 &  \n &2073.84 &2023.64 \\ 
0003$-$066  & 2 & 5  & 222 &1.05&  5.12& $ 322.9$ &
 226.3$\pm$4.8 & 96.6$\pm$5.0\tablenotemark{c} & 191$\pm$15 & 4.09$\pm$0.33 &  \n &1993.56 &2004.48 \\ 
  & 3 & 9  & 119 &2.82&  13.73& $ 296.9$ &
 284.8$\pm$4.5 & 12.1$\pm$4.6 & 250$\pm$39 & 5.36$\pm$0.83 &  \n &1989.50 &1979.99 \\ 
  & 4\tablenotemark{a} & 26  & 119 &6.61&  32.23& $ 285.6$ &
 283.6$\pm$8.1 & 1.9$\pm$8.1 & 41$\pm$13 & 0.87$\pm$0.28 &  \n &1845.03 &1820.94 \\ 
  & 5 & 6  & 1031 &0.73&  3.58& $ 14.5$ &
 343.1$\pm$3.1 & 31.4$\pm$3.1\tablenotemark{c} & 100$\pm$16 & 2.15$\pm$0.35 &  \n &2011.53 &1997.84 \\ 
  & 6\tablenotemark{a} & 10  & 97 &1.01&  4.92& $ 290.2$ &
 210$\pm$13 & 81$\pm$13\tablenotemark{c} & 55$\pm$17 & 1.18$\pm$0.37 &  \n &1969.25 &2011.24
\enddata 
\tablenotetext{a}{Feature has significant accelerated motion (see Table~\ref{acceltable} for acceleration fit parameters).}
\tablenotetext{b}{Feature has significant inward motion.}
\tablenotetext{c}{Feature has significant non-radial motion.}
\tablenotetext{d}{Fit parameters for this source are from \cite{MOJAVE_X}.}
\tablenotetext{e}{Feature has slow pattern speed.}
\tablenotetext{~}{A question mark indicates a feature whose motion is not consistent with outward, radial motion but for which the possibility of inward motion and its degree of non-radialness are uncertain.}
\tablecomments{This is a table stub, the full version is available as an ancillary file. 
Columns are as follows: (1) B1950 name, (2) feature number, (3) number of fitted epochs, (4) mean flux density at 15 GHz in mJy,  (5) mean distance from core feature in mas, (6) mean projected distance from core feature in pc, (7) mean position angle with respect to the core feature in degrees, (8) position angle of velocity vector in degrees, (9) offset between mean position angle and velocity vector position angle in degrees,  (10) angular proper motion in microarcseconds per year, (11) fitted speed in units of the speed of light, (12) estimated epoch of origin, (13) fitted epoch of origin in right ascension direction, (14) fitted epoch of origin in declination direction.}
\end{deluxetable*} 

\begin{deluxetable*}{lrrrrrrrrrr} 
\tablecolumns{11} 
\tabletypesize{\scriptsize} 
\tablewidth{0pt}  
\tablecaption{\label{acceltable}Acceleration Fit Properties of Jet Features}  
\tablehead{\colhead{} & \colhead {}& \colhead{$\phi$} &    \colhead{$ |\langle\vartheta\rangle - \phi|$}& 
\colhead{$\mu$} & \colhead{$\beta_{app}$} & \colhead{$ \dot{\mu}_{\perp} $}  & \colhead{$ \dot{\mu}_{\parallel} $} && & \\  
\colhead{Source} & \colhead {I.D.} & \colhead{(deg)} & \colhead{(deg)} &   
\colhead{($\mu$as y$^{-1})$}&\colhead{(c)} &\colhead{($\mu$as y$^{-2})$}&\colhead{($\mu$as y$^{-2})$}&\colhead{$ t_\mathrm{mid}$}  & \colhead{$t_\alpha$}& \colhead{$t_\delta$}   \\  
\colhead{(1)} & \colhead{(2)} & \colhead{(3)} & \colhead{(4)} &  
\colhead{(5)} & \colhead{(6)} & \colhead{(7)}  & \colhead{(8)} & \colhead{(9)}  & \colhead{(10)}& \colhead{(11)}  } 
\startdata 
0003+380  & 6  & $ 333 \pm 40 $ & $ 142 \pm 40 $& $ 13.4 \pm 8.5 $ & 0.20$\pm$0.12 &  $ -4 \pm 11$ $(\pm 8.3)$ & $ 9.0 \pm 9.2$ $(\pm 7.6)$  & 2009.90 & 2071.12 & 2024.51\\ 
0003$-$066  & 4\tablenotemark{a}  & $ 270.0 \pm 4.1 $ & $ 15.6 \pm 4.1\tablenotemark{b} $& $ 50.4 \pm 5.3 $ & 1.08$\pm$0.11 &  $ 8.5 \pm 2.5$ $(\pm 1.6)$ & $ -27.4 \pm 2.4$ $(\pm 2.3)$  & 2004.83 & 1874.35 & 503767\\ 
  & 6\tablenotemark{a}  & $ 211.3 \pm 8.8 $ & $ 78.9 \pm 8.9\tablenotemark{b} $& $ 54 \pm 11 $ & 1.16$\pm$0.24 &  $ 54 \pm 13$ $(\pm 11)$ & $ -37 \pm 17$ $(\pm 15)$  & 2003.78 & 1971.75 & 2009.60\\ 
  & 8\tablenotemark{a}  & $ 290.7 \pm 1.6 $ & $ 3.5 \pm 1.7 $& $ 330.4 \pm 9.8 $ & 7.08$\pm$0.21 &  $ -19 \pm 12$ $(\pm 12)$ & $ -64 \pm 12$ $(\pm 12)$  & 2009.93 & 2002.02 & 2000.32\\ 
  & 9  & $ 295.2 \pm 4.1 $ & $ 7.5 \pm 4.2 $& $ 278 \pm 20 $ & 5.96$\pm$0.42 &  $ 9 \pm 36$ $(\pm 36)$ & $ -99 \pm 35$ $(\pm 35)$  & 2009.24 & 2002.22 & 2004.34\\ 
0010+405  & 1  & $ 340.7 \pm 4.3 $ & $ 11.9 \pm 4.3 $& $ 432 \pm 42 $ & 6.99$\pm$0.68 &  $ 11 \pm 54$ $(\pm 53)$ & $ -43 \pm 69$ $(\pm 69)$  & 2008.52 & 1978.76 & 1991.39\\ 
  & 2  & $ 359 \pm 237 $ & $ 31 \pm 237 $& $ 3 \pm 17 $ & 0.04$\pm$0.28 &  $ 2 \pm 16$ $(\pm 16)$ & $ -3 \pm 26$ $(\pm 26)$  & 2008.87 & $-$37937 & 1427.76\\ 
  & 3  & $ 138 \pm 94 $ & $ 170 \pm 94 $& $ 2.6 \pm 4.6 $ & 0.043$\pm$0.074 &  $ -4.3 \pm 9.6$ $(\pm 6.6)$ & $ 5.4 \pm 9.8$ $(\pm 7.0)$  & 2008.87 & 2290.10 & 2404.54\\ 
  & 4  & $ 114 \pm 190 $ & $ 146 \pm 190 $& $ 1.4 \pm 3.5 $ & 0.023$\pm$0.057 &  $ -2.1 \pm 8.8$ $(\pm 7.2)$ & $ -4.7 \pm 7.5$ $(\pm 5.4)$  & 2008.87 & 2195.70 & 2676.67\\ 
0016+731  & 1\tablenotemark{a}  & $ 163.2 \pm 2.2 $ & $ 22.8 \pm 2.4\tablenotemark{b} $& $ 106.2 \pm 4.4 $ & 8.23$\pm$0.34 &  $ 10.2 \pm 1.8$ $(\pm 1.8)$ & $ 8.8 \pm 2.0$ $(\pm 1.9)$  & 2002.70 & 1975.08 & 1994.50
\enddata 
\tablenotetext{a}{Feature shows significant accelerated motion.}
\tablenotetext{b}{Feature shows significant non-radial motion.}
\tablenotetext{c}{Feature shows significant inward motion.}
\tablenotetext{d}{Fit parameters for this source are from \cite{MOJAVE_X}.}
\tablenotetext{~}{A question mark indicates a feature whose motion is not consistent with outward, radial motion but for which the possibility of inward motion and its degree of non-radialness are uncertain.}
\tablecomments{This is a table stub, the full version is available as an ancillary file. 
Columns are as follows: (1) B1950 name, (2) feature number, (3) proper motion position angle in degrees,  (4) offset between mean position angle and velocity vector position angle in degrees, (5) angular proper motion in microarcseconds per year, (6) fitted speed in units of the speed of light, (7) angular acceleration perpendicular to velocity direction in microarcseconds per year per year, with the error excluding the direction uncertainty in parentheses, (8) angular acceleration parallel to velocity direction in microarcseconds per year per year,  with the error excluding the direction uncertainty in parentheses, (9) date of reference (middle) epoch used for fit, (10) fitted epoch of origin in right ascension direction, (11) fitted epoch of origin in declination direction.}
\end{deluxetable*} 

\begin{figure*}[p]
\centering
\includegraphics[trim=0cm 0cm 0cm 0cm,height=0.88\textheight]{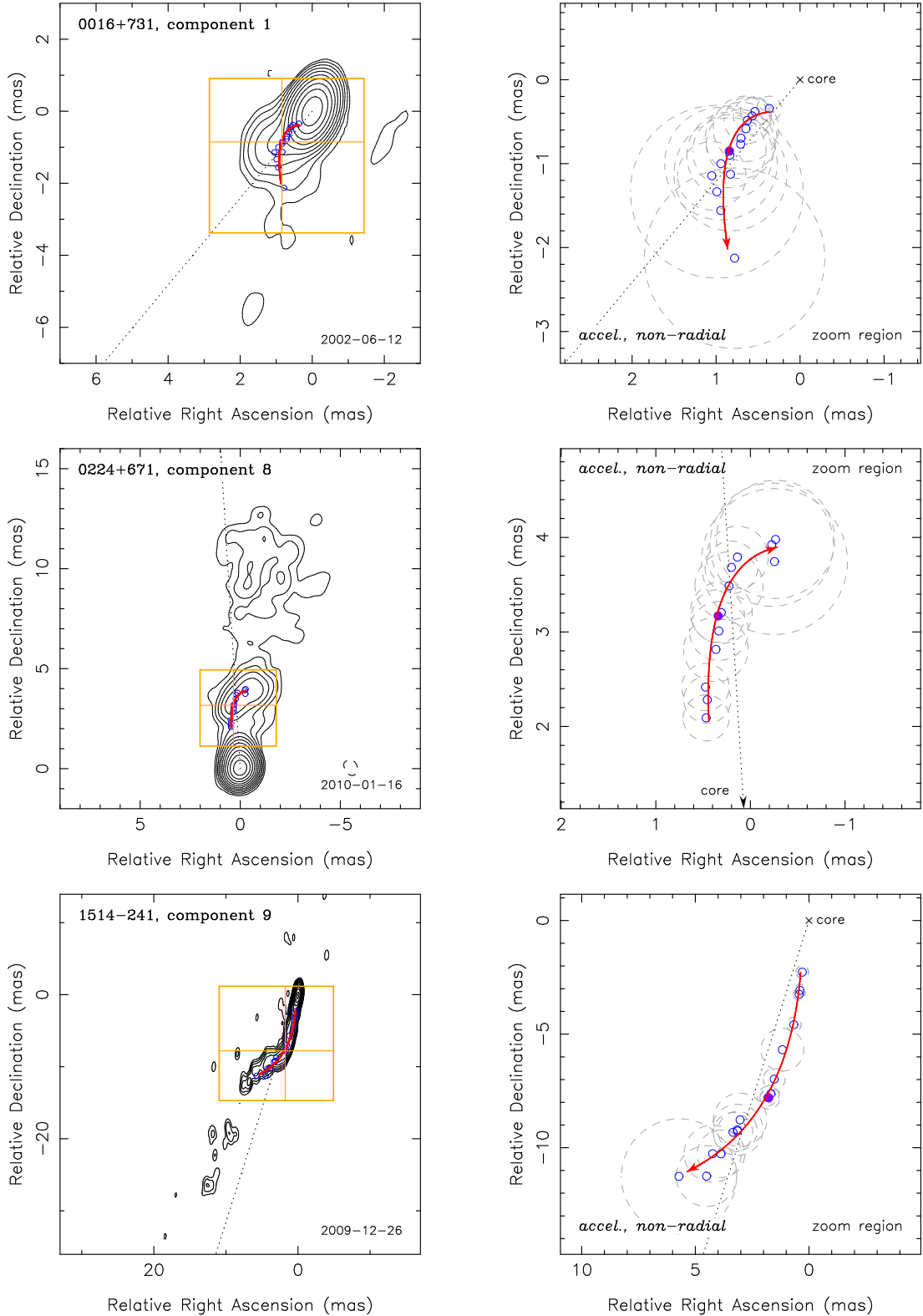}
\caption{\label{xyplot_a}
  Vector motion fits and sky
  position plots of individual robust jet features in MOJAVE AGNs.
  Positions are relative to the core position. The left hand panels
  show a 15 GHz VLBA contour image of the jet at the epoch listed in
  Table~\ref{acceltable}, which is closest to the (midpoint) reference
  epoch.  The orange box delimits the zoomed region that is displayed
  in the right hand panels. The feature's position at the image epoch
  is indicated by the orange cross-hairs.  The dotted line connects
  the feature with the core feature and is plotted with the mean
  position angle $\langle\vartheta\rangle$. The position at the image
  epoch is shown by a filled violet circle while other epochs are
  plotted with unfilled blue circles. The red solid line indicates the
  vector fit (or accelerating fit, if there is significant
  acceleration) to the feature positions. The gray dashed
  circles/ellipses represent the FWHM sizes of the individual fitted
  Gaussian features. A red vector motion direction arrow is indicated
  in all of the plots, including features for which the angular speed
  is extremely slow and/or statistically insignificant. This is a
  figure stub. The full set is available online at
  \url{http://www.astro.purdue.edu/MOJAVE/velocitytable.html}.}
\end{figure*}

In the left hand panels of Figure~\ref{xyplot_a} we show the positions
and kinematic motion fits to the individual features on the sky, as
well as a 15 GHz VLBA contour image of the jet at the observation
epoch that is closest to the (midpoint) reference date.  The orange
box delimits the zoomed region displayed in the right hand panels,
with the cross-hairs indicating the feature's position at this epoch.

\begin{figure}
\centering
\includegraphics[trim=0cm 0cm 0cm 0cm,angle=270,width=0.47\textwidth]{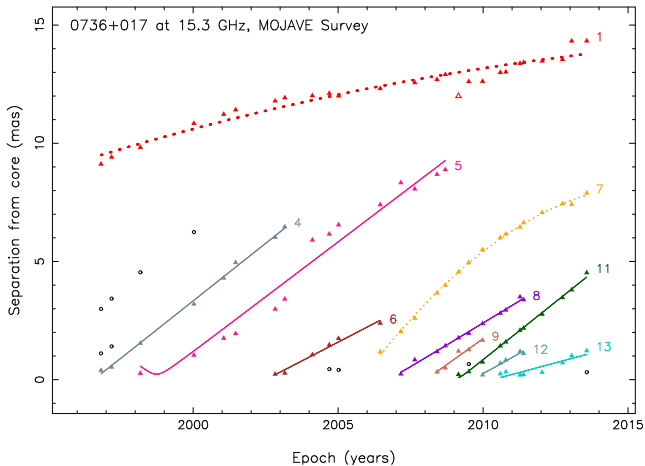}
\caption{\label{sepvstime} 
  Plot of angular separation from
  the core versus epoch for Gaussian jet features. The B1950 source
  name is given at the top left of each panel. Colored symbols
  indicate robust features for which kinematic fits were obtained
  (dotted and solid lines). The solid lines indicate vector motion
  fits to the data points assuming no acceleration, while the dotted
  lines indicate accelerated motion fits. Thick lines are used for
  features whose fitted motion is along a radial direction from the
  core, while thin lines indicate non-radial motions. Unfilled colored
  circles indicate individual data points that were not used in the
  kinematic fits, and unfilled black circles indicate non-robust
  features.  The feature identification number is indicated next to
  the last epoch of each robust feature. 
  This is a figure stub, the full set of plots is available at \url{http://www.astro.purdue.edu/MOJAVE/allsources.html}.}
\end{figure}

We note that many of the jet features in our sample display non-radial
motion, i.e., the fitted velocity vector at the reference epoch does
not point back to the core feature.  This is indicative of
accelerated, non-ballistic motion.  In some cases, an extrapolation of
our simple constant acceleration fit shows the feature passing through
the core feature at the ejection epoch within the errors, as expected.
However, in many other cases, the acceleration(s) are either complex,
occurred before our monitoring period, or took place close to the core
below our angular resolution level.  These unknown accelerations
strictly limit the number of features for which we can derive reliable
ejection epochs, and can cause an apparent inflection in the fit line
at minimum separation (e.g., feature id = 5 of 0736$+$017) in the
plots of radial separation from the core versus time (see
Figure~\ref{sepvstime}).

We examined each feature with significant proper motion $(\mu \ge
3\sigma_\mu)$ for non-radial motion by comparing the mean position
angle of the feature $\langle\vartheta\rangle$ with its proper motion
vector direction $\phi$.  We flagged any feature for which the angular
offset $|\langle\vartheta\rangle - \phi|$ was $\ge3\sigma$ from 
both $0\arcdeg$ and $180\arcdeg$ as non-radial, and inward if the offset was
significantly greater than 90\arcdeg.  Of the 714 newly analyzed
features with statistically significant ($\ge3\sigma$) proper motion,
228 (32\%) exhibit significant non-radial motion, while only 25 (4\%)
are flagged as inward (see Table~\ref{velocitytable}). The latter can
be the result of curved trajectories that are approaching the line of
sight, internal brightness changes in a large, diffuse jet feature, or
backward pattern speeds not associated with the flow. We discuss these
rare individual inward motion cases in the Appendix.

We calculated ejection times (epoch of zero core separation) for the
non-accelerating, non-inward moving features with significant motion
by taking an error-based weighted average of the extrapolated
emergence epochs in the right ascension ($t_\alpha$) and declination
($t_\delta$) directions (columns 13 and 14 of
Table~\ref{velocitytable}). The individual values of $t_\alpha$ and
$t_\delta$ describe the fits, but do not necessarily have physical
meaning, depending on the proper motion vector direction, the
statistical significance of the fitted motion, and/or accelerations
that may be present.  We include them in the motion tables for
completeness only.  We used the method described in \cite{MOJAVE_X} to
calculate the errors on the ejection times.

Due to the nature of our proper motion fits, which naturally include
the possibility of non-ballistic motion, we did not estimate ejection
times for any features where we could not confidently extrapolate
their motion to the core.  Jet features for which we estimated
ejection epochs had the following properties: (i) significant motion
$(\mu \ge 3\sigma_\mu)$, (ii) no significant acceleration, (iii) a
velocity vector within $15\arcdeg$ of the outward radial direction to
high confidence, i.e., $|\langle\vartheta\rangle - \phi|+2\sigma \le
15\arcdeg$, (iv) an extrapolated position at the ejection epoch no
more than $0.2$ mas from the core, and (v) a fitted ejection epoch
that differed by no more than 0.5 years from that given by a simple
one-dimensional (radial) motion fit.

We list the parameters of our acceleration fits in
Table~\ref{acceltable}, where we have resolved the acceleration terms
$\dot\mu_\perp$ and $\dot\mu_\parallel$ in directions perpendicular
and parallel, respectively, to the mean angular velocity direction
$\phi$ (columns 7 and 8). We quote two uncertainty values in these
acceleration terms.  The value in parentheses is the statistical
uncertainty due to the acceleration fit itself, and is the appropriate
quantity for determining if a significant acceleration occurred
relative to the velocity vector direction.  The total uncertainty
value incorporates an estimate of the effect of a possibly poorly
known velocity direction when assigning these accelerations the labels
``parallel'' or ``perpendicular''.  For example, the uncertainty on
the parallel acceleration adds in quadrature the term $\sqrt{\left(
    \dot\mu_\perp d\phi\right)^2 + \left((\dot\mu_\parallel
    d\phi)d\phi\right)^2}$, where $d\phi$ is the uncertainty in the
velocity direction in radians, and takes on a maximum value of 1.0 in
this expression. An analogous term is added in quadrature to the
statistical uncertainty for the perpendicular acceleration.  In our
previous papers, we listed the acceleration fit uncertainty alone, and
did not include any additional error due to uncertainty in the vector
motion direction. Instead, as described in \cite{MOJAVE_VII} and
\cite{MOJAVE_XII}, we limited our detailed acceleration analysis in
those papers to jet features where the velocity direction was very
well known and therefore the parallel and perpendicular assignments
were unambiguous.

\begin{figure*}
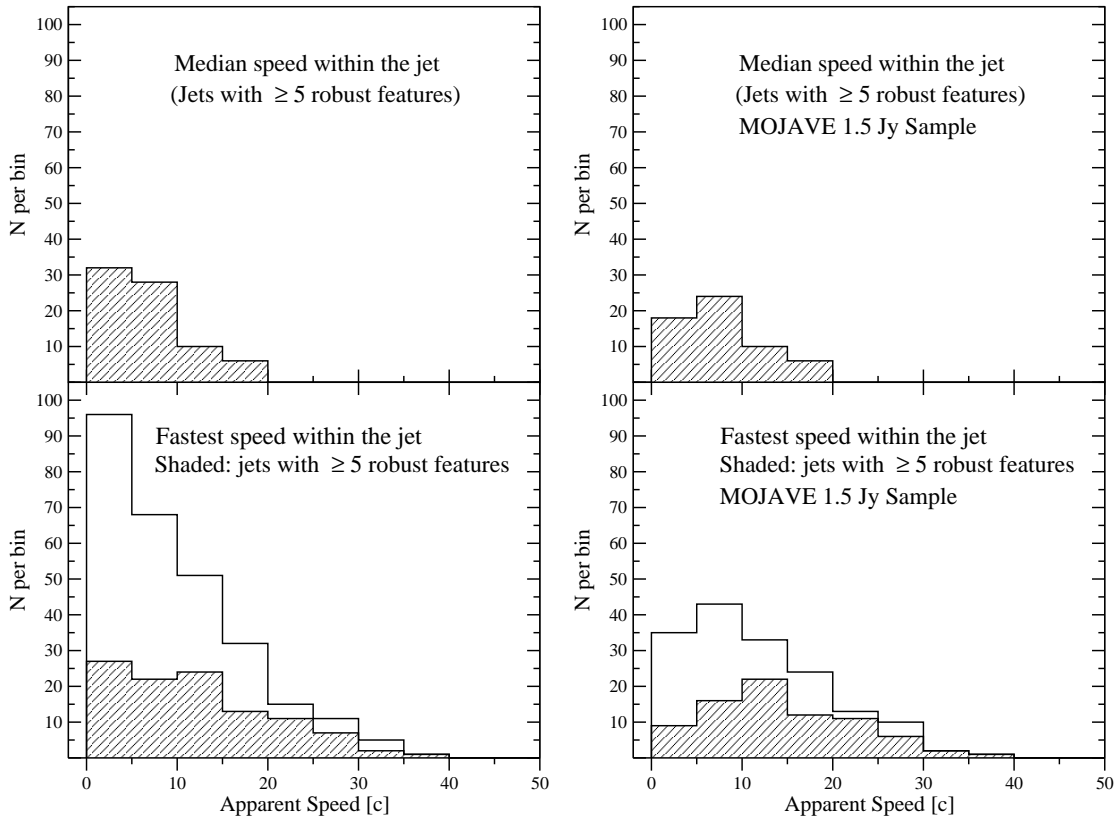

\centering
\includegraphics[trim=0cm 0cm 0cm 0cm,clip,width=0.4\textwidth]{betaspeedhist.eps}
~
\includegraphics[trim=0cm 0cm 0cm 0cm,clip,width=0.4\textwidth]{betaspeedhist_mojave.eps}
\caption{\label{speedhist} 
  Distributions of median  and
  fastest speeds for the jets in our sample with measured speeds and redshifts.  The
  shaded histograms include only those jets with at least 5 robust features.
  The left-hand panels contain all speeds for all AGNs measured by the MOJAVE
  program to date, while the right-hand panels are for AGNs in the complete
  flux density-limited 1.5 Jy MOJAVE sample only.}
\end{figure*}

\begin{figure*}
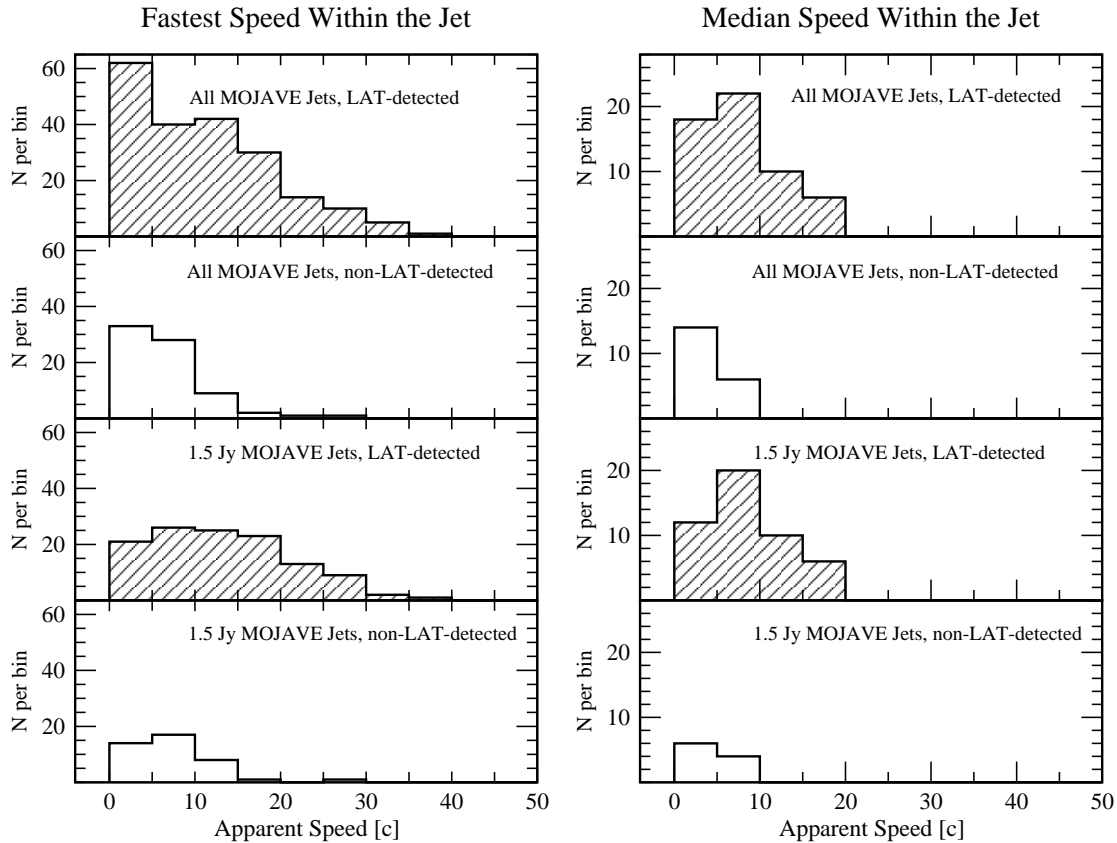

\centering
\includegraphics[trim=0cm 0cm 0cm 0cm,clip,width=0.4\textwidth]{betaspeedhist_LAT.eps}
~
\includegraphics[trim=0cm 0cm 0cm 0cm,clip,width=0.4\textwidth]{betamedianspeedhist_LAT.eps}
\caption{\label{LATspeedhist}
  Jet speed distributions for
  various MOJAVE sub-samples. The left-hand panels show the
  distributions for the fastest measured jet speed, while the right
  hand panels show the distributions of median jet speed for jets
  having at least 5 robust features. The topmost two panels show all
  speeds measured by the MOJAVE program to date for {\it Fermi} LAT
  detected (shaded) and non-LAT detected AGN (unshaded), respectively.
  The bottom two panels show the speeds of the LAT detected (shaded)
  and non-LAT detected (unshaded) AGN, respectively, in the flux
  density-limited 1.5 Jy MOJAVE sample.}
\end{figure*}

Of the 451 features in our current analysis with more than 10 epochs,
144 (32\%) show significant accelerations, and an additional 87 show
non-radial motions. Thus over half of the well-sampled jet features in
our survey show evidence of non-ballistic motion. In agreement with
our previous findings \citep{MOJAVE_VII,MOJAVE_XII}, apparent speed
changes are larger and more prevalent than changes in direction,
indicating that changes in Lorentz factor are more responsible for the
observed accelerations than jet bends.

\section{\label{discussion}DISCUSSION}

\subsection{Jet Speed Distributions}

We have calculated maximum and median speed statistics for all 274 AGN
jets measured in this paper, and 48 AGNs in \cite{MOJAVE_X}, using the
method described in the latter paper. For those AGNs that were
analyzed in both papers, we used the kinematic results from the
current paper. The combined set of 322 AGNs spans the full sky above
declination $-30^\circ$, and all sources have at least $\sim 0.1$ Jy
of VLBA correlated flux density at 15 GHz.  Although the set is
complete above 1.5 Jy \citep{2015ApJ...810L...9L} and at high
$\gamma$-ray (time-averaged) fluxes \citep{2011ApJ...742...27L}, the
selection is incomplete and non-uniform at lower radio flux densities,
containing a sampling of gamma-ray-associated and un-associated AGNs,
as well as lower-luminosity AGN targeted specifically to probe the
relation between jet speed and radio luminosity \citep{MOJAVE_X}.
Because of the $>0.1$ Jy VLBA limit at 15 GHz, the combined set of AGN has
a substantial bias toward core-dominated, flat-spectrum radio sources,
which are expected to be highly beamed.  This is reflected in the
optical classifications, which are dominated by blazars (i.e., flat
spectrum radio quasars: 73\% and BL Lac objects: 18\%). There are also
22 radio galaxies and 5 narrow-lined Seyfert I (NLSY1) galaxies, and 4
AGNs with no optically identified counterparts.  A total of 243 AGNs
have published gamma-ray associations, which are listed in column 4 of
Table~\ref{gentable}.

In Figure~\ref{speedhist} we plot the speed distributions for the combined
set of AGNs (left hand panels), and for the radio flux density-limited
MOJAVE 1.5 Jy sample only (right hand panels).  As discussed by \cite{VC94}
and \cite{LM97}, the shape of the apparent speed distribution in a
flux density-limited jet sample reflects the parent Lorentz factor
distribution, which in this case, indicates a moderately steep parent
power law, with fast jets being relatively rare. A more detailed
analysis of the parent population properties will be presented in a
future paper.  Because of Doppler bias, the fastest apparent speeds in
a large flux density-limited sample should also be indicative of the
maximum bulk flow Lorentz factors in the parent population. The
fastest overall jet speed we have measured in our survey to date is
feature id = 2 in the quasar PKS 0805$-$07, with $\mu = 503 \pm 62
\muasyr$, corresponding to $40c \pm 5c$.  In terms of angular
expansion rate, the maximum values of our survey are limited to
roughly $<3000 \muasyr$ due to our limited angular resolution and
observing cadence.  The fastest measured angular speed in our survey
to date is $2941\pm 109\muasyr$, in the jet of the nearby ($z
=0.033$) AGN 3C~120 \citep{MOJAVE_X}.

We note that there are two accelerating features whose apparent speed
exceeded $40c$ at some point during our monitoring (the speeds listed
in Table~5 are the instantaneous speeds at the midpoint epoch). These
are feature id = 3 in PKS 0805$-$07, which had an initial speed of
$\sim 50c$ in 1996 and subsequently decelerated to $\sim 20c$, and
feature id = 16 in PKS 1510$-$08, which attained a maximum speed of
$\sim 42c$ before fading in 2011.

In \cite{MOJAVE_X}, we identified 38 jet features as having slow
pattern speeds, i.e., they had non-accelerating, angular speeds
smaller than 20 \muasyr, and a speed at least 10 times slower than the
fastest feature in the same jet. With the addition of our new
kinematics analysis, we now have identified a total of 63 such
features (5\% of all robust features), which we have flagged in
Table~\ref{velocitytable}.

\subsection{Jet Speed and Gamma-Ray Emission}

The most recent catalog of gamma-ray sources detected by the {\it
  Fermi} LAT instrument (3FGL: \citealt{3FGL}) is highly dominated by
blazars, due to the strong influence of relativistic beaming on jet
gamma-ray emission. As discussed by \cite{2015ApJ...810L...9L}, based
on data from \cite{MOJAVE_X}, nearly all of the fastest AGN jets in
the MOJAVE program have been detected by {\it Fermi}, indicating a
strong correlation between jet speed and gamma-ray Doppler boosting
factor. Several fast jets have yet to be detected by {\it Fermi}
because their spectral energy distribution peaks below the 100 MeV
threshold of the LAT.  In Figure \ref{LATspeedhist}, we show the
distributions of fastest and median jet speed for our full MOJAVE
sample (top panels), and our 1.5 Jy flux density-limited sample (lower
panels), including our new speed data.  In terms of the fastest
measured speeds, there is a less than 0.02 \% probability that the LAT
and non-LAT sub-samples come from the same parent distribution
according to Kolomogorov-Smirnov tests, in the case of both the full
and MOJAVE 1.5 Jy flux density-limited samples.





\begin{figure*}
\centering
\includegraphics[trim=0cm 0cm 0cm 0cm,angle=270,width=0.8\textwidth]{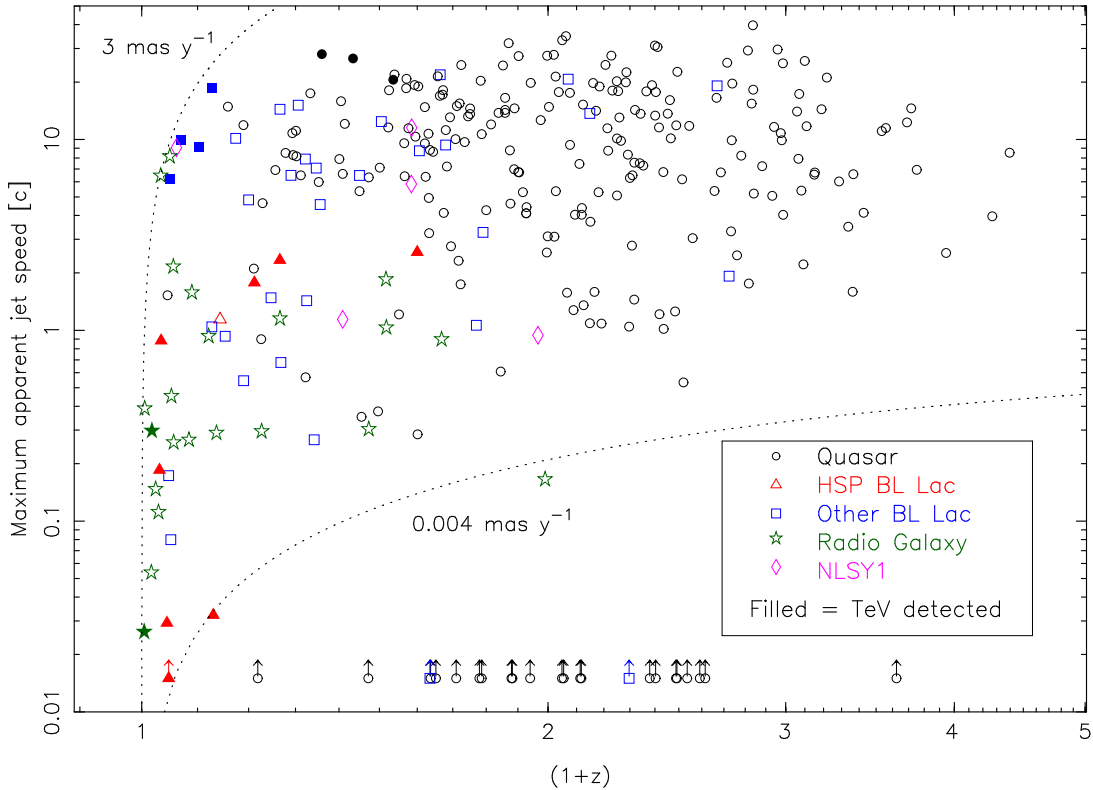}
\caption{\label{logbapp_vs_logz}
  Log-log plot of maximum apparent
  jet speed versus redshift for all jets measured by the MOJAVE
  program to date. The dotted lines delimit the approximate survey
  limits on angular proper motions, based on our angular resolution,
  temporal coverage, and observing cadence.  Black circles correspond to
  quasars, red triangles to high spectral peaked BL~Lac objects, blue squares
  to other BL Lacs, green stars to radio galaxies, and purple diamonds to
  NLSY1s.  Filled symbols indicate AGNs that have been detected at TeV
  gamma-ray energies.  Jets for which were not able to identify any
  robust features for kinematic analysis are plotted as lower limits
  with an arbitrary value of 0.015~c.}
\end{figure*}

\subsection{Apparent Jet Speed Versus Redshift}

In Figure~\ref{logbapp_vs_logz} we plot maximum measured apparent jet
speed versus redshift for all jets measured by the MOJAVE program to
date.  For completeness, we also plot as lower limits equal to $0.015c$
those AGN jets for which we were not able to identify any robust
features for kinematics analysis, despite having at least 5 VLBA
epochs.

This collection of AGNs represents those selected from the MOJAVE
radio flux density-limited, gamma-ray-selected, and low-luminosity
blazar samples.  As such, it spans a very large range of redshift
($0.004 < z < 3.4$) and apparent radio luminosity ($10^{23}-10^{29}\;
\mathrm{W\; Hz^{-1}}$). At low redshifts, lower luminosity radio
galaxies and BL Lac objects dominate, since high luminosity quasars
are much rarer in the general AGN population (see, e.g., discussion of
luminosity functions in \citealt{UP95}).

Our monthly observing cadence, total time baseline, and angular
resolution set limits on the maximum and minimum observable angular
proper motions, which we plot as dotted lines in
Figure~\ref{logbapp_vs_logz}.  We note that for $z > 0.3$, our survey
is sensitive to extremely high jet speeds, yet there is a
distinct upper envelope to the distribution. Since there must be a
portion of the ultrafast AGN jet population that is oriented close to
our line of sight, and such jets will be extremely bright due to
enormous Doppler boosting of their radio flux density, the upper
envelope in Figure~\ref{logbapp_vs_logz} is a clear indication of an
upper limit of $\sim 50$ in the bulk Lorentz factors in the
pc-scale radio-emitting regions of AGN jets.

In Figure~\ref{logbapp_vs_logz} we have also indicated with filled
symbols those AGNs that have been detected at TeV gamma-ray energies.
To date, there are roughly 60 known AGNs in this
category\footnote{\url{http://tevcat.uchicago.edu}}, and one third of these
have been monitored by the MOJAVE program. Our plot illustrates the
two types of known TeV AGN: i) BL Lac objects that have high TeV
fluxes due to their proximity and the fact that their spectral energy
distributions are peaked at high frequencies; and ii) other BL Lac
objects and quasars with very fast jet speeds, indicating unusually
high amounts of relativistic beaming.  To date there have been no TeV
detections of AGNs past $z \simeq 1$ due to the strong attenuation of
high energy gamma-rays by pair production off of the extragalactic
background light. This would suggest that a promising region to search
for new TeV-emitting AGNs is in the top left corner of
Figure~\ref{logbapp_vs_logz}, where $\beta_\mathrm{app} > 10c$ and $z
< 0.35$. This region includes the BL Lac objects PKS~0754+100, OJ~287
and OJ~049, the quasars 3C~273, OI~061, 4C~+49.22, TXS 1700+685, and
PKS~1725+044, the AGNs 3C~111 and 3C~120, and the NLSY1 1H~0323+342.

We have measured maximum jet speeds for 9 high spectral peaked (HSP)
BL Lacs to date in the MOJAVE program (triangle symbols in
Figure~\ref{logbapp_vs_logz}), all but one of which (7C 1055+5644)
have been detected at TeV energies. Their apparent speeds are
sub-luminal, ranging from $0.03c \pm 0.01c$ in ON~325 to $2.6c \pm
 1.1c$ in OQ~240. This is in agreement with previous multi-epoch
VLBI observations of a small number of TeV blazar jets at 8 GHz by
\cite{2008ApJ...678...64P}.  We are currently obtaining multi-epoch
MOJAVE observations of a much larger set of high-spectral peaked AGN
detected by the {\it Fermi} $\gamma$-ray observatory in order to
further investigate their speed distribution.

\subsection{Radio-loud Narrow-Lined Seyfert I AGNs}

The third {\it Fermi} AGN catalog \citep{3LAC} is comprised primarily of
blazars, reflecting the strong influence of relativistic boosting on
AGN gamma-ray emission. Apart from a small handful of nearby radio
galaxies, the only other major AGN category in the {\it Fermi} catalog
is radio-loud NLSY1 galaxies. To date fewer than a dozen
gamma-ray-loud NLSY1s have been identified \citep{2015arXiv151200171B},
but they are interesting objects since they appear to have many of the
same characteristics as blazars, despite the fact that Seyferts are
typically found in spiral host galaxies. The orientation of NLSY1s
renders their broad line region visible, yet they display narrow
permitted lines ($< 2000\; \mathrm{km\; s^{-1}}$) due to their small
black hole masses ($10^6 - 10^8 M_\sun$). Therefore in order to
produce blazar-like gamma-ray and radio luminosities, they must have
high Eddington ratios.  This has led to speculation that they might be
young radio jets that have been recently fueled, and are perhaps the
low-black-hole mass, low-viewing angle versions of compact symmetric AGNs (e.g.,
\citealt{2009ApJ...693.1686K}).

We have analyzed the pc-scale jet kinematics of five NLSY1s in the
MOJAVE sample, all of which have been detected in gamma-rays by {\it
  Fermi}.  Other groups (\citealt{2012MNRAS.426..317D,
  2013MNRAS.433..952D}) have published simple one-dimensional
radial-fit speeds for two of these AGNs using our MOJAVE data.  We
find that three of the NLSY1s show highly superluminal jet features
($9.0c \pm 0.3c$ for 1H 0323+342, $5.8c \pm 0.9c$ for SBS 0846+513,
and $11.5c \pm 1.5c$ for PMN J0948+0022). The other two show
kinematics consistent with sub-luminal speeds: $0.9c \pm 0.3c$ for 4C
+04.42 and $1.1c \pm 0.4c$ for 1502+036.  With this small sample size,
it is impossible to draw any general conclusions about the population,
but the high speeds indicate that at least some radio-loud NLSY1s have
Lorentz factors in excess of 10, and viewing angles less than
$10\arcdeg$, consistent with those of BL Lac objects and flat-spectrum
radio quasars.

\section{SUMMARY}\label{conclusions}

We have presented 1625 new 15 GHz VLBA contour images of 295 AGNs from
the MOJAVE and 2cm VLBA surveys, spanning epochs from 1994 Aug 31 to
2013 Aug 20. For the 274 AGNs with at least 5 temporally-spaced
epochs, we have analyzed the kinematics of individual bright features
in the jets, and produced time-lapse linearly interpolated movies
that show their parsec-scale evolution. At least half of the
well-sampled features show non-radial and/or accelerating
trajectories, indicating that non-ballistic motion is common. Since it
is impossible to extrapolate any accelerations that occurred before
our monitoring period or below our resolution level, we are only able
to determine reliable ejection dates for $\sim 24\%$ of those features with
significant proper motions.

The distribution of maximum jet speeds in all 295 AGNs measured by our
program to date is peaked below $5c$, with very few jets with apparent
speeds above $30c$.  The fastest instantaneous measured speed in our
survey is $\sim 50c$, measured in the jet of the quasar PKS~0805$-$07.
The form of the speed distribution is indicative of a moderately steep
power law distribution of Lorentz factors in the parent population,
ranging up to $\Gamma \simeq 50$.  An upper envelope in the maximum
speed versus redshift distribution provides additional evidence of
this upper limit to AGN jet speeds in the pc-scale regime.

The {\it Fermi} LAT-detected gamma-ray AGNs in our sample have, on
average, higher jet speeds than non LAT-detected AGNs, indicating a
strong correlation between pc-scale jet speed and gamma-ray Doppler
boosting factor.  The known TeV gamma-ray emitting AGNs in our sample
are clustered in two regions: namely at low redshift/low speeds, and
at moderate redshift/high speeds. We have identified 11 AGNs in the
latter region ($\beta_\mathrm{app}> 10c$, $z < 0.35$) that are strong
candidates for future TeV gamma-ray detection. Our AGN sample also
includes 5 gamma-ray-loud NLSY1s, a rare type of low-black hole mass,
high-accretion rate AGN that has been identified as a distinct
population of gamma-ray emitting AGNs by {\it Fermi}. Of the five
NLSY1s, three show highly superluminal jet motions, while the others
have sub-luminal speeds.  This indicates that radio-loud NLSY1 AGNs
can possess powerful jets with high Lorentz factors and low viewing
angles that are consistent with typical radio loud BL Lac objects and
flat-spectrum radio quasars.

\acknowledgments
The MOJAVE project was supported by NASA-{\it Fermi} GI grants
NNX08AV67G, NNX12A087G, and NNX15AU76G.  The National Radio Astronomy
Observatory is a facility of the National Science Foundation operated
under cooperative agreement by Associated Universities, Inc.  E.R.
acknowledges partial support by the the Spanish MINECO project
AYA2012-38491-C02-01 and by the Generalitat Valenciana project
PROMETEOII/2014/057.  MFA was supported in part by NASA-{\it Fermi} GI
grants NNX09AU16G, NNX10AP16G and NNX11AO13G, NNX13AP18G and NSF grant
AST-0607523.  YYK and ABP were supported by the Russian Foundation for
Basic Research (project 13-02-12103) and the Basic Research Program
P-7 of the Presidium of the Russian Academy of Sciences.  TS was
supported by the Academy of Finland projects 274477 and 284495.  This
work made use of the Swinburne University of Technology software
correlator \citep{2011PASP..123..275D}, developed as part of the
Australian Major National Research Facilities Programme and operated
under licence.

{\it Facilities:} \facility{VLBA}.


\appendix{\bf{}Notes on Individual AGNs}

\begin{myparindent}{0pt}

  0006+061 (CRATES J0009+0628): \cite{2012AA...538A..26R} obtained $z
  < 1.35$ for this BL Lac based on photometric redshift estimates. It
  is one of two possible 3LAC associations for 3FGL J0009.1+0630
  listed by \cite{3LAC}. The other is CRATES J0009+0625.

  0048$-$097 (PKS 0048$-$09): The jet structure is too compact at 15
  GHz to reliably measure any robust features.

  0106+678 (4C +67.04): This BL Lac has an inward-moving feature (id
  = 4) located 1.3 mas from the core.

  0109+224 (S2 0109+22) : The redshift $z = 0.265$ from
  \cite{2012ApJ...748...49S} is claimed to be incorrect
  according to \cite{2016arXiv160208703P}, who find a lower limit of
  $z > 0.35$, based on new optical spectroscopy.

  0108+388 (CGRABS J0111+3906): Multi-frequency VLBA images of this
  radio galaxy indicate a two-sided jet morphology, however, the
  precise location of the core is not known
  \citep{2001ApJ...550..160M}. We therefore used the southwestern-most
  bright feature in the images as a reference point for the
  kinematics analysis.

  0111+021 (UGC 00773): With the addition of one more epoch in 2013
  since our \cite{MOJAVE_X} analysis, we now classify a total of four features in this BL
  Lac jet as inward-moving.

  0118$-$272 (OC $-$230.4): This BL Lac has a very slow inward-moving
  feature ($25 \pm 7 \muasyr$) located 0.5 mas from the core.
  \cite{2013ApJ...764..135S} found  $z <$ 0.558 based on intergalactic
  absorption features in the optical spectrum.

  0119+041 (PKS 0119+041): The jet structure is too compact at 15 GHz
  to reliably measure any robust features.

  0141+268 (TXS 0141+268): \cite{2009ApJ...704..477S} reported $z <$
  2.41 for this BL Lac object.

  0219+428 (3C 66A): The two innermost jet features of this BL Lac
  object have robust inward motions. No reliable spectroscopic
  redshift has been published to date. \cite{2014ApJ...784..151S} reported $z
  >$ 0.42 based on the host galaxy magnitude.  

  0235+164 (AO 0235+164): The jet structure is too compact at 15 GHz
  to reliably measure any robust features.

  0300+470 (4C +47.08): No reliable spectroscopic redshift exists for
  this BL Lac is unknown \cite{2013ApJ...764..135S}, which contains an
  inward-moving feature (id = 1) at 0.8 mas from the core feature.
 
  0301$-$243 (PKS 0301$-$243): The innermost feature of this BL Lac
  jet has a robust inward motion.

  0346+800 (S5 0346+80): No reliable spectroscopic redshift exists in
  the literature for this BL Lac object.

  0414$-$189 (PKS 0414$-$189): The jet structure is too compact at 15
  GHz to reliably measure any robust features.

  0440$-$003 (NRAO 190): This quasar is one of two possible 3LAC
  associations for 3FGL J0442.6$-$0017 listed by \cite{3LAC}.  The
  other is a radio-quiet X-ray source (1RXS J044229.8$-$001823) which
  is 2.3 arcmin from the LAT position (as compared to 0.21 arcmin for
  0440$-$003).

  0454+844 (S5 0454+84): The value of $z = 1.340$ from
  \cite{2001AJ....122..565R} is a lower limit, and the value $z = 0.113$
  from \cite{1996ApJS..107..541L} is considered tentative.

  0506+056 (TXS 0506+056): As described by \cite{MOJAVE_X}, this BL
  Lac object has an inward-moving feature (id = 3) at 1.3 mas from the
  core feature.

  0615+820 (S5 0615+82): The jet structure is too compact to reliably
  identify the core position at all epochs, or to classify any jet
  features as robust.

  0640+090 (PMN J0643+0857): The position of the core feature in
  this low-galactic latitude quasar is uncertain, and we considered
  none of the jet features to be robust. We used the highest
  brightness temperature feature as a reference point for the
  kinematics analysis.

  0646+600 (S4 0646+60): This is a high-frequency peaked quasar
  \citep{2000AA...363..887D} with two-sided parsec-scale radio
  structure, meeting the criteria for a compact symmetric object.
  The outermost jet feature at $\sim 3$ mas from the core has slow but
  significant inward motion ($10 \pm 0.7 \muasyr$).

  0727$-$115 (PKS 0727$-$11): The radio morphology of this quasar was
  too compact and complex to identify any suitably robust jet
  features for kinematic analysis.


  0742+103 (J0745+1011): This high redshift quasar has a
  gigahertz-peaked spectrum and jet emission to both the NW and SE of
  the brightest feature in the maps. The core location is unknown,
  making it impossible to reliably cross-register the maps across the
  epochs or to classify any jet features as robust.

  0743$-$006 (J0745-0044): The position of the core in this
  gigahertz-peaked spectrum AGN is uncertain. We used the position
  of the most compact feature in the maps as a reference point for
  the kinematics analysis.

 0745+241 (S3 0745+24): The jet feature at 3.4 mas from the core (id
 = 11) has significant inward motion.

 0804+499 (OJ 508): This quasar is one of two possible 3LAC
 associations for 3FGL J0807.9+4946 listed by \cite{3LAC}. The other
 is a double-lobed radio galaxy (J0807+4946) with no visible radio
 core in the FIRST survey image, located 7.74 arcmin from the LAT
 position (as compared to 0.77 arcmin for OJ~508).

 0814+425 (OJ 425): The value of $z = 0.53$ from
 \cite{2012ApJ...744..177L} is uncertain, while
 \cite{2013ApJ...764..135S} reported $z <$ 2.47 and
 \cite{2005ApJ...635..173S} obtained $z >$ 0.75. As described in
 \cite{MOJAVE_X}, the innermost feature (id = 5) has robust inward
 motion.

 0821+394 (4C +39.23): This quasar is one of two possible 3LAC
 associations for 3FGL J0824.9+3916 listed by \cite{3LAC}. The other
 is the compact steep-spectrum quasar 4C +39.23B
 \citep{2002AA...389..115D}.

 0823$-$223 (PKS 0823$-$223): The value of $z = 0.91$ from
 \cite{1990ApJ...353..114F} is a lower limit.  We could not identify
 any suitably robust jet features for kinematic analysis in this BL
 Lac.

 0946+006 (PMN J0948+0022): Using additional VLBA epochs on this
 narrow-lined Seyfert I galaxy obtained since our \cite{MOJAVE_X}
 analysis, we now classify two jet features as robust.

 0954+658 (S4 0954+65): The widely cited redshift of $z = 0.367$ by
 \cite{1993AAS...98..393S} has been drawn in to question by
 \cite{2015AJ....150..181L}, who find a featureless spectrum and a
 limit $z > 0.45$ based on non-detection of the host galaxy. We note
 that for $z > 1.7$, our apparent speed measurement of $\mu = 672 \pm
 50 \muasyr$ would yield $v_\mathrm{app} \simeq 50c$, a value
 larger than the fastest measured speed of any AGN in the MOJAVE
 program.

 1030+415 (S4 1030+41): The jet structure is too compact at 15 GHz to
 reliably measure any robust features.

 1101+384 (Mrk 421): The innermost feature (id = 8) in this nearby
 BL Lac jet has slow but significant inward motion.


 1124$-$186 (PKS 1124$-$186): The jet structure is too compact at 15
 GHz to reliably measure any robust features.

 1128+385 (B2 1128+38): The outermost feature in this jet has
 significant inward motion.

 1144+402 (S4 1144+40): This quasar is one of two possible
 associations listed for 3FGL J1146.8+3958 by \cite{3LAC}. The other
 is the weak radio source NVSS J114653+395751, which showed less than
 1 mJy of correlated VLBA flux density at 1.4 GHz in the mJIVE-20
 survey of \cite{2014AJ....147...14D}.

 1148$-$001 (4C $-$00.47): The position of the core in this high
 redshift ($z = 1.98$) quasar is uncertain. We used the position of
 the most compact feature in the maps as a reference point for the
 kinematics analysis.

 1213$-$172 (PKS 1213$-$17): This AGN is located very close on the sky
 to a bright star, and has no reliable spectroscopic redshift in the
 literature.

 1219+044 (4C +04.42): \cite{2015MNRAS.454L..16Y} have classified this
 AGN as a NLSY1 galaxy, based on its optical spectrum from the
 SDSS-BOSS survey.

 1219+285 (W Comae): A feature (id = 17) located $\sim 1$ mas from the core
 has significant inward motion.

 1243$-$072 (PKS 1243$-$072): This quasar is listed as an association
 in the third EGRET catalog \citep{Hartman99}, but not in any of the
 {\it Fermi} catalogs.

 1324+224 (B2 1324+22): The jet structure is too compact at 15 GHz to
 reliably measure any robust features.

 1329$-$126 (PMN J1332$-$1256): We did not identify any suitably
 robust jet features for kinematic analysis in this quasar.

 1331+170 (OP 151): This quasar is listed as an association in the
 third EGRET catalog \citep{Hartman99}, but was not confirmed in the
 revised EGRET catalog of \cite{2008AA...489..849C}. It is not listed
 in any of the {\it Fermi} catalogs.

 1413+135 (J1415+1320): This unusual BL Lac object resides in a spiral
 host galaxy \citep{2002AJ....124.2401P}, and has a two-sided jet
 structure \citep{2005ApJ...622..136G}. The  core is associated
 with the brightest feature in our VLBA images.

 1435+638 (VIPS 0792): We presumed the radio core to lie at the
 northernmost end of the jet, based on the VIPS 5 GHz VLBA map of
 \cite{2007ApJ...658..203H}.

 1458+718 (3C 309.1): This compact steep spectrum quasar has a group
 of inward moving jet features located 23 mas south of the core. On
 larger scales, the radio morphology suggests a twisting (helical)
 jet structure \citep{1998MNRAS.299..467L}.

 1508$-$055 (PKS 1508$-$05): This compact steep spectrum quasar is
 listed as a gamma-ray association in the 3LAC catalog \citep{3LAC}.
 We find no significant proper motion  in two robust jet features.

 1509+054 (PMN J1511+0518): This gigahertz-peaked spectrum radio
 galaxy has a two-sided jet morphology \citep{MOJAVE_XI}, and a
 statistically significant inward moving feature (id = 2). However,
 it is possible that we are merely seeing changes in the internal
 brightness distribution of unresolved features in this jet, and not
 true apparent motion.

 1510$-$089 (PKS 1510$-$08): Many of the features in this jet (e.g.,
 id = 8,15,16) show complex kinematics, including a kink in their
 trajectories at 2 mas from the core that are not well approximated by
 a constant acceleration model.

 1514+004 (PKS 1514+00): The brightness profile along this nearby ($z
 = 0.052$) radio galaxy jet is very smooth, and not well-represented
 by discrete Gaussian features.

 1519$-$273 (PKS 1519$-$273): The jet structure is too compact at 15
 GHz to reliably measure any robust features.

 1529$-$131 (PMN J1532$-$1319): There is no known optical counterpart
 for this AGN.  We could not identify any suitably robust jet
 features for kinematic analysis.

 1548+056 (4C +05.64): The southernmost jet feature, which we presume
 marks the location of the core, faded in the 2005 October 29 epoch,
 thus we have not used this epoch in our kinematics analysis. 

 1611+343 (DA 406): The morphology and complex kinematics of this
 quasar jet suggest that it is bending into our line of sight several
 milliarcseconds from the core, and that we may be seeing features
 from both the front and back sides of the jet.

 1622$-$253 (PKS 1622$-$253): The jet structure is too compact at 15
 GHz to reliably measure any robust features.

 1637+826 (NGC 6251): In \cite{MOJAVE_X} we reported an inward-moving
 feature (id = 8) in this radio galaxy jet. With the addition of two
 new epochs, this feature no longer shows any statistically
 significant motion.

 1656+053 (PKS 1656+053): We could not identify any suitably robust
 jet features for kinematic analysis in this quasar. 

 1656+477 (S4 1656+47): We could not identify any suitably robust jet
 features for kinematic analysis in this quasar.

 1739+522 (4C +51.37): The jet structure is too compact at 15 GHz to
 reliably measure any robust features.

 1741$-$038 (PKS 1741$-$03): The jet structure is too compact at 15
 GHz to reliably measure any robust features.

 1749+701 (S4 1749+70 : All of the jet features in this BL Lac show
 transverse motions, including the innermost feature, which evolves
 from a position angle of $-50\arcdeg$ to $-75\arcdeg$ over a 15 year
 period. 

 1842+681 (S4 1842+68): We could not identify any suitably robust jet
 features for kinematic analysis in this quasar.

 1923+210 (PKS B1923+210): This low galactic latitude AGN has a
 featureless optical spectrum, as reported by
 \cite{2011AJ....142..165T}.

 1958$-$179 (PKS 1958$-$179): The jet structure is too compact at 15
 GHz to reliably measure any robust features.

 2021+614 (OW 637): The core feature location in this jet remains
 uncertain \citep{MOJAVE_VI}.  The two outermost features (ID = 1
 and 2) have inward motions, albeit with very slow speeds (8.8 and
 19.6 $\muasyr$, respectively).

 2023+335 (B2 2023+33): This low galactic latitude quasar is strongly
 affected by interstellar scattering \citep{2013AA...555A..80P}. The
 radio structure was too complex to reliably cross-identify any
 robust features for kinematic analysis. Examples of this kind of
 strong scattering are extremely rare among bright, compact AGN
 \cite{2015MNRAS.452.4274P}. 

 2047+098 (PKS 2047+098): This AGN has no known optical counterpart.
 The value $z = 0.01513$ listed in the 2LAC \citep{2LAC} data table is
 unreferenced. Its jet has two inward moving features (id = 2 and
 3).

 2128$-$123 (PKS 2128$-$12): The overall kinematics and polarization
 structure of this quasar underwent sudden and dramatic changes in
 2009.

 2144+092 (PKS 2144+092): We could not identify any suitably robust
 jet features for kinematic analysis in this quasar.

 2155+312 (B2 2155+31): We could not identify any suitably robust jet
 features for kinematic analysis in this quasar.

 2230+114 (CTA 102): By using additional VLBA epochs obtained since
 \cite{MOJAVE_X}, we now identify two robust inward-moving features (id = 1
 and 4) in this quasar jet. These were also identified in the
 multi-frequency VLBI kinematic analysis of this AGN by
 \cite{2013AA...551A..32F}.

 2234+282 (CTD 135): \cite{2016arXiv160103859A} have claimed this BL
 Lac to be a compact symmetric object with a two-sided jet, based on a
 comparison of VLBA maps at 8.4 and 15 GHz. However, the angular
 resolution is insufficient to determine a spectral index for the
 putative core feature. For the purposes of our kinematic analysis we
 have identified the core with the northernmost jet feature in our
 maps.

 2247$-$283 (PMN J2250$-$2806): By using additional VLBA epochs we
 have obtained since \cite{MOJAVE_X}, we are able to identify one robust feature
 in this quasar jet.

 2356+196 (PKS 2356+196): This quasar is listed as an association in
 the third EGRET catalog \citep{Hartman99}, but was not confirmed in
 the revised EGRET catalog of \cite{2008AA...489..849C}. It is not
 listed in any of the {\it Fermi} catalogs.

\end{myparindent}

\medskip

\end{document}